\documentclass[letter, twocolumn]{jpsj2} 

\title{Liquid-Solid Transition and Phase Diagram of $^4$He Confined in Nanoporous Glass}

\author{Keiichi \textsc{Yamamoto}\thanks{E-mail address: kyamamot@phys.keio.ac.jp}, 
Yoshiyuki \textsc{Shibayama},
Keiya \textsc{Shirahama}
}

\inst{Department of Physics, Keio University, Yokohama 223-8522, Japan}

\abst{
We have studied the liquid - solid (L-S) phase transition of $^4$He confined in nanoporous 
glass, which has interconnected nanopores of 2.5~nm in diameter.
The L-S boundary is determined by the measurements of pressure and thermal response during slow cooling and warming. 
Below 1 K, the freezing pressure is elevated to 1.2 MPa from the bulk freezing pressure, and appears to be independent of temperature.
The $T$-independent L-S boundary implies the existence of a localized Bose-Einstein condensation state, 
in which long-range superfluid coherence is destroyed by narrowness of the nanopores and random potential.
}

\kword{Superfluidity, Bose-Einstein Condensation, Quantum Phase Transition, Porous Media}

\begin{document}
\maketitle


Quantum phase transition (QPT) has been of great interest in condensed matter physics~\cite{SondhiSachdev}. 
Bose systems in periodic or random potential offers a typical example for studies of QPT. 
QPT in ultracold atomic gases~\cite{Greiner2002}, thin superconducting films~\cite{Merchant2001}, 
Josephson junction arrays~\cite{Geerligs1989}, and high-temperature superconductors~\cite{Jiang1994} have been extensively studied.
These systems undergo transitions from superconducting (superfluid) to insulating (localized) states such as a Mott insulator or a Bose glass~\cite{Fisher1989}.

Bose systems under potential are not only important for understanding QPT, 
but also provide us with novel aspects in the 
physics of Bose-Einstein condensation (BEC) and superfluidity. 
$^4$He in porous media is an important model system as strongly correlated Bosons in external potential.
One can control freely many properties such as the dimensionality, the topology, and the disorder of the system.
The degree of controllability 
is unique and unavailable in the other systems. 

For the last three decades, a number of experimental studies have been carried out for $^4$He in porous media~\cite{Reppy1984, Reppy2000,Shirahama1990,Wada2001,Chan1988}. 
Effects of disorder on critical phenomena of superfluid $^4$He were investigated using Aerogel~\cite{Chan1988}. 
In porous Vycor glass, a dilute Bose gas state was demonstrated \cite{Reppy2000}.
These studies were done in low-density adsorbed film states and liquid states at ambient pressure.
In contrast, little attention has been given to confined $^4$He \textit{under pressure}.
One may expect novel quantum phenomena in  pressurized $^4$He in restricted geometries, which is an interesting correlated Bose system.

\begin{figure}[t]
\centering
    \includegraphics[width=70mm,keepaspectratio]{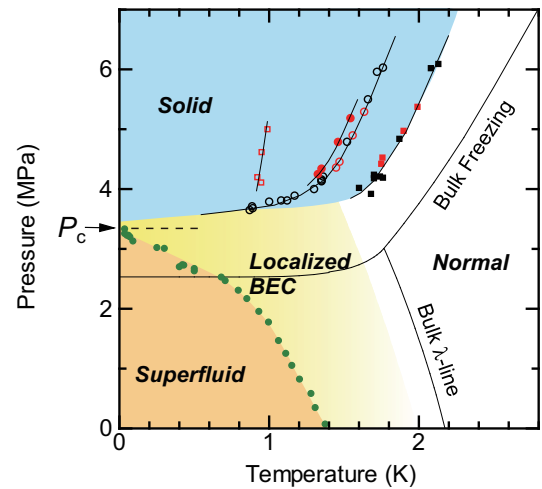}
  \caption{\label{PhaseDiagram}
  (Color online)
  The $P$-$T$ phase diagram of $^4$He in the 2.5~nm nanoporous glass determined by our present and previous works.
  The green data points show the superfluid $T_{\mathrm c}$ obtained by the torsional oscillator study \cite{Yamamoto2004}. 
  The black and red data points are taken by Cell 1 and Cell 2, respectively.
  The open and closed circles indicate the freezing onset and completion, respectively.
  The open and closed squares  show the melting onset and completion, respectively.
  The solid lines show the bulk $\lambda$ line and the bulk L-S boundary.
}
\end{figure}

Recently, we studied the superfluid transition of $^4$He 
confined in a nanoporous Gelsil glass of 2.5~nm pore diameter using a torsional oscillator technique~\cite{Yamamoto2004}.
We summarize the result in Fig.~\ref{PhaseDiagram} together with the results of the present work.
The confinement drastically suppresses superfluidity:  
 the superfluid transition temperature $T_\mathrm{c}$ is about 1.4~K at saturated vapor pressure,
which is already suppressed to 2/3 of the bulk superfluid transition temperature ($T_\lambda$).
As pressure increases, $T_\mathrm{c}$  dramatically decreases and approaches 0~K at a critical pressure $P_\mathrm{c}~=~3.4~\mathrm{MPa}$.
In addition, the superfluid density also decreases continuously to zero as pressure approaches $P_\mathrm{c}$~\cite{Yamamoto2004}.
These unprecedented behaviors clearly show that the confined $^4$He undergoes a QPT at the quantum critical pressure $P_\mathrm{c}$.


The liquid-solid (L-S) transition of $^4$He in the nanopores is expected to occur far above the bulk freezing pressure.
The existence of the quantum critical pressure shows that a nonsuperfluid (NSF) phase exists at 0~K.
If the NSF phase were the real normal fluid, the third law of thermodynamics had to be violated.
The L-S phase boundary will provide us with the information on the thermodynamics of the NSF phase.
Torsional oscillator technique cannot distinguish the normal liquid from the solid phase, 
because the torsional oscillation responds to not only a solid but also a normal viscous liquid in which the viscous penetration depth is much larger than the pore size.
In this paper,  we report on the L-S transition of $^4$He in the nanopores by the measurements of pressure and thermal response, 
in the temperature range from 0.02~K to 2.5~K and for pressures up to 6~MPa.
A preliminary result has been reported elsewhere~\cite{Yamamoto2006}.            

We have measured a number of pressure-temperature ($P-T$) isochores using a
low-temperature pressure gauge attached to the sample cell wall~\cite{Straty1969}.
The sample cell is made of BeCu, and contains a stack of three or four porous glass disks.
The pressure is obtained from the change in the capacitance through the deflection of the cell wall which acts as a diaphragm.
In the sample cell, the porous glasses are inevitably surrounded by bulk solid $^4$He  in the gap between the glasses and the cell.
This causes poor pressure transmission from $^4$He in the nanopores to the diaphragm wall.
To reduce the pressure transmission time, we have set the gap between the glass samples and the diaphragm to be 0.4~mm. 
The temperature of the cell is measured by a Ge and a RuO$_2$ bare chip resistors.
They were calibrated against a calibrated Ge thermometer by Lake Shore Cryotronics Inc.

We have carried out the pressure measurements with two sample cells.
The first cell, referred to as Cell 1, was mounted directly on a mixing chamber of a dilution refrigerator.
Some signatures of the L-S transition were observed in the pressure of Cell 1. 
However, no thermal anomaly was observed, 
because it was difficult to sweep the temperature  smoothly  above 1.2~K, at which the dilution refrigerator is not in stable operation.
We have constructed the second cell, Cell 2. 
This cell is mounted on a massive Cu isothermal stage by a support tube which is made from a rolled Kapton sheet of 25~$\mu$m thick. 
The cell is thermally linked to the stage by eight NbTi superconducting wires of $130~\mu {\mathrm m}$ in diameter,   
which originally act as wirings to the thermometers. 
The external time constant of the cell for thermal equilibrium is approximately 5-10 times the relaxation time for equilibrium within the cell.
This long external time constant enables us to control the cell temperature precisely and measure the thermal response in very slow temperature sweeps.
The isothermal stage is mounted and weakly thermal-linked to the mixing chamber.
Because of the weak thermal links, the achievable lowest temperature of Cell 2 was limited to 400~mK, whereas Cell 1 was cooled down to 20~mK.

\begin{figure}[t]
\centering
    \includegraphics[width=70mm, keepaspectratio]{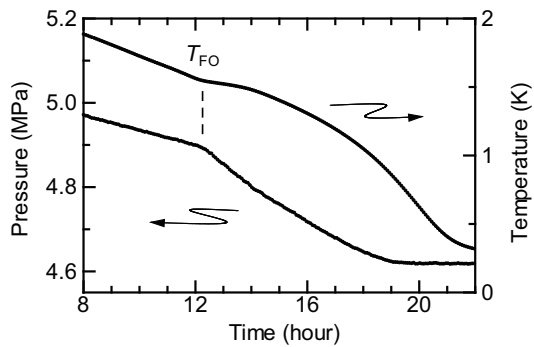}
\caption{\label{TimeDependence} 
Temperature and pressure in cooling by Cell 2, as a function of time. The initial pressure $P_\mathrm{ini}$ is $8.1~\mathrm{MPa}$.
}
\end{figure}

In Cell 1 and 2, we employed three and four Gelsil disks, respectively. 
Each disk sample is 5.5~mm in diameter and 2.3~mm in height, and is taken from the same batch as that employed in our previous torsional oscillator experiment.  
The surface area of one disk is obtained to be 26.9~m$^2$ by a N$_\mathrm{2}$ adsorption measurement at 77~K.
Before mounting, the samples are heated at 150 Ž in vacuum for 3 hours at 2$\times $10$^{-4}$~Pa to remove adsorbed molecules.

Since liquid $^4$He is usually supercooled inside the nanopores, 
the porous glass sample is surrounded by bulk solid $^4$He in the measurement of the L-S transition. 
We prepare the bulk solid $^4$He by the blocked capillary method.
After feeding liquid $^4$He into the cell using a room-temperature gas handling system,
the sample cell is initially  pressurized to $P_\mathrm{ini}$ ranging from 5 to 8.5~MPa at 4.2~K, 
and then is cooled slowly by operating the dilution refrigerator.
Solidification proceeds from the filling capillary above the mixing chamber to the sample cell.
The pressure of the cell decreases along the bulk L-S boundary on the $P-T$ plane.
After the entire bulk $^4$He in the cell solidifies, the pressure separates from the  L-S boundary.
In the cooling-warming runs of Cell 1, we observed pressure hysteresis just below the freezing curve, because of the poor crystallinity of the freshly grown solid. In Cell 2, the solid sample was annealed just below the bulk freezing temperature for 10~hours. This eliminated the hysteresis.
The cooling and warming speeds of Cell 2 are  1.5-4.5~mK/min.

Figure~\ref{TimeDependence} shows the typical data of cooling Cell 2, starting from $P_\mathrm{ini}=8.1~\mathrm{MPa}$.
$P$ drops abruptly at 
1.55~K, which is referred to as $T_\mathrm{FO}$. 
Simultaneously, the cooling rate decreases.
The pressure drop at $T_\mathrm{FO}$ is attributed to the onset of freezing of $^4$He in the nanopores.
The decrease in the cooling rate indicates the release of the latent heat of freezing.

\begin{figure}[t]
\centering
    \includegraphics[width=75mm, keepaspectratio]{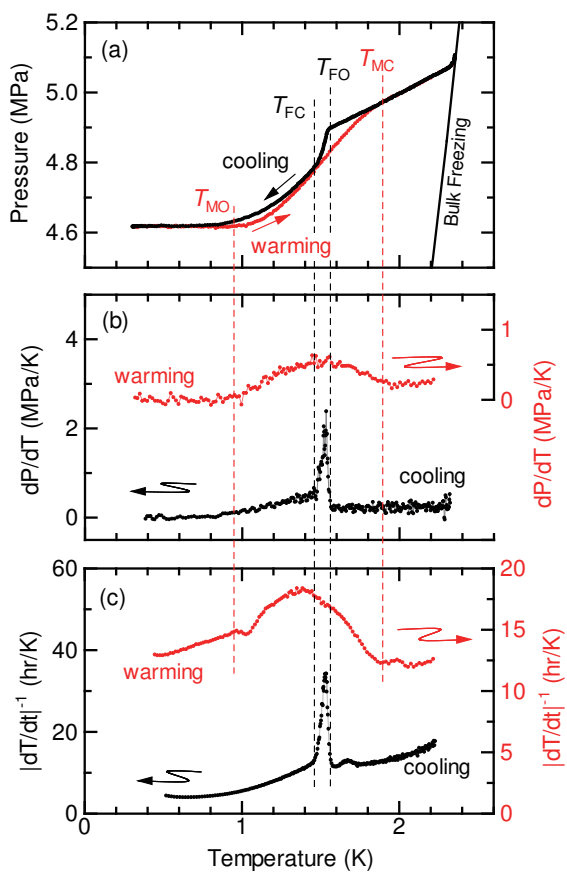}
\caption{\label{P-T} 
(Color online)
(a) The pressure isochores: 
Black line is cooling data, which are replotted from the time dependences of temperature and pressure shown in Fig~\ref{TimeDependence}.
Red line is taken at warming.
(b) Corresponding $dP/dT$.
(c) Inverse of $dT/dt$. 
Four vertical lines indicate, 
$T_\mathrm{FO}$: freezing onset temperature in the cooling run, 
$T_\mathrm{FC}$: freezing completion,
$T_\mathrm{MO}$: melting onset in the warming run,
$T_\mathrm{MC}$: melting completion, respectively.
}
\end{figure}

The behavior of $P(T)$ corresponding to Fig.~\ref{TimeDependence} is shown in Fig.~\ref{P-T}(a).
Below the onset temperature $T_\mathrm{FO}$, the abrupt pressure drop proceeds in a finite temperature range of approximately 100 mK.
Below 1~K, pressure becomes independent of temperature down to the lowest temperature (20~mK in Cell 1, and 400~mK in Cell 2).
We have confirmed that $P(T)$ is taken in nearly thermal equilibrium, 
from the fact that several isochores taken with different cooling rates collapse onto the same curve.
All the observations in $P(T)$ are reproduced in Cell~1.
In the warming run (from 400~mK in Cell 2, and 20~mK in Cell 1), the isochore traces the cooling path up to 
0.9~K.
In the temperature range between 0.9~K and 1.9~K, the isochore shows hysteresis. 
Similar behavior was observed in the case of $^4$He 
confined in Vycor and other porous glasses~\cite{Adams1984, Adams1987, Beamish1983, Bittner1994}.

To show the thermal and pressure anomalies more clearly, 
we show the temperature derivative of the pressure $dP/dT$ and the time derivative of the temperature $|dT/dt|^{-1}$ of Cell~2 as a function of temperature 
in Fig.~\ref{P-T}(b) and (c).
In the present setup of Cell 2, 
the temperature change of the cell is caused by a heat current $\dot{Q}$ from the isothermal stage to the cell.
In this case $dT/dt$ is related to the total heat capacity of the cell $C_\mathrm{tot}$ and $\dot{Q}$ by 
\begin{equation}
\dot{Q}\left( T \right) = C_\mathrm{tot} \left( dT / dt \right).
\label{Qdot}
\end{equation}
Although $\dot{Q}$ is not measured by the present setup, it is reasonable to assume that $\dot{Q}$ is a smooth function of temperature in the range of measurement.
Therefore, anomalies in $C_\mathrm{tot}$ originated from  phase transitions are detected as signatures in $|dT/dt|^{-1}$.

In the cooling run, both $dP/dT$ and  $|dT/dt|^{-1}$ start to increase at $T_\mathrm{FO}$ and show single peaks.
Both peaks terminate at $T_\mathrm{FC}$, which we assign to the temperature of freezing completion.  
The peak structure in $|dT/dt|^{-1}$ corresponds to the heat capacity peak caused by the latent heat of freezing.
The $|dT/dt|^{-1}$ peak is sharp, but rounded unlike the first-order L-S transition in bulk system.
The confinement into the narrow pores and the pore-size distribution inherent to the porous glass
may produce the broadening of the L-S transition.

In the warming process, both  $dP/dT$ and $|dT/dt|^{-1}$ show peak structures as well.
The peaks are, however, much broader than those in the cooling. 
The broad $|dT/dt|^{-1}$ peak is attributed to the heat absorption caused by the melting of $^4$He in the nanopores.
In Fig.~\ref{P-T}(b) and (c), $dP/dT$ and  $|dT/dt|^{-1}$ start to increase simultaneously at $T_\mathrm{MO}$.
We assign $T_\mathrm{MO}$ to the temperature of melting onset.
The $P(T)$ curves on cooling and warming collapse onto a single curve above a temperature shown as $T_\mathrm{MC}$.
Above $T_\mathrm{MC}$ both  $dP/dT$ and  $|dT/dt|^{-1}$ stop to decrease.
We conclude that the melting is completed at $T_\mathrm{MC}$.
\begin{figure}[t]
\centering
    \includegraphics[width=65mm, keepaspectratio]{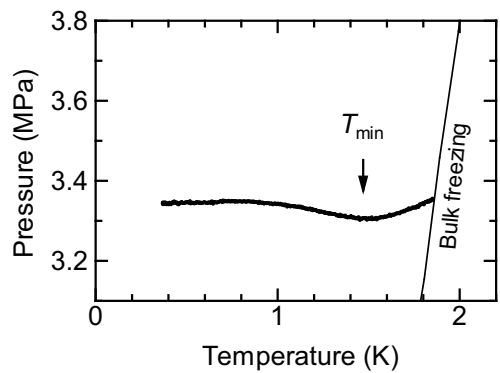}
\caption{\label{PMinimum} 
Pressure data at low initial pressure $P_\mathrm{ini} = 6.53~\mathrm{MPa}$ measured by Cell 2.
A pressure minimum is observed at a temperature denoted as $T_\mathrm{min}$.
}
\end{figure}

We have performed the measurements for various starting pressures. 
The pressure drop at $T_\mathrm{FO}$ in the cooling run becomes smeared as $P_\mathrm{ini}$ decreases.
The minimum freezing pressure that we have observed was $P_\mathrm{FO} = 3.7~\mathrm{MPa}$ at $T_\mathrm{FO} = 0.8~\mathrm{K}$.
The freezing curve was not determined below 3.7~MPa due to its weak temperature dependence.
However, we conclude that the freezing curve is located within the narrow pressure range of $3.4 < P < 3.7~\mathrm{MPa}$.
The pressure and the thermal response shows no indications of freezing at all up to 3.35~MPa, as shown in Fig~\ref{PMinimum}. 
In the sample of Cell 1,  no freezing was observed up to 3.4~MPa and down to 20~mK.
Hence, below 3.4~MPa, $^4$He in the nanopores remains supercooled liquid down to at least 20~mK.
In this pressure range where the isochores have no signature of freezing, 
 $P(T)$ shows a minimum at a temperature denoted as $T_\mathrm{min}$ as shown in Fig.~\ref{PMinimum}.
We will comment on this behavior later.

In Fig.~\ref{PhaseDiagram}, we show the freezing and melting curves of $^4$He in Gelsil, 
 with $T_\mathrm{c}$ of $^4$He in Gelsil and the bulk phase boundaries.
The  curves shifts to lower temperatures and to higher pressures. 
The elevation of the freezing pressure $\Delta P$ from that of bulk is found to be 1.43~MPa at 1.3~K.
Similar elevation was observed in various $^4$He systems confined in porous media.
Adams \textit{et al.} studied the overpressure of $^4$He in the 7~nm Vycor glass and in a porous Bioglass which has 2.4~nm pores~\cite{Adams1984,Adams1987, Bittner1994}. 
The observed $\Delta P$ for the 7~nm and 2.4~nm glasses were 1.16 and 1.40~MPa at 1.3~K, respectively.
$\Delta P$ in our 2.5~nm Gelsil glass is slightly higher than that of 2.4~nm-Bioglass data.

The freezing pressure elevation in porous media has been interpreted in terms of the homogeneous nucleation theory~\cite{Beamish1983, Adams1984,Adams1987, Bittner1994}. 
Because solid $^4$He does not wet the substrate of porous glass, solid nucleation from the substrate is suppressed.
Solidification must therefore take place in the liquid apart from the glass substrate.
$\Delta P$ to form a homogeneous solid droplet of radius $R$ 
is given by 
\begin{equation}
\Delta P = \frac{2 \alpha _\mathrm{LS} v_\mathrm{S}}{R(v_\mathrm{L}-v_\mathrm{S})},
\label{deltaP}
\end{equation}
where $\alpha _\mathrm{LS}$ is the L-S interface tension, and $v_\mathrm{S}$ and $v_\mathrm{L}$ are the molar volume of the solid and liquid, respectively.
Because the size of the solid droplet cannot exceed the pore size, this equation gives the elevated pressure for freezing.
In the present study, $\alpha _\mathrm{LS}$ is estimated to be 0.09~erg/cm$^2$ using the obtained value $\Delta P = 1.43~\mathrm{MPa}$ at 1.3~K
and assuming that $v_\mathrm{S}$ and $v_\mathrm{L}$ are same as those of bulk $^4$He.
This value is comparable to the liquid - hcp solid interfacial tension of $^4$He at temperatures below 1.5~K, 0.16~erg/cm$^2$ \cite{Landau1980}.
Although it is not clear if the assumption on the molar volumes in the nanopores is correct,
the homogeneous nucleation theory can account semi quantitatively for the freezing pressure elevation in the nanopores.


The phase diagram of $^4$He in Gelsil shows that  a nonsuperfluid (NSF) phase exists between the
superfluid and solid phase down to 0 K.
Below 1 K, the freezing pressure has little temperature dependence.
According to the Clapeyron-Clausius relation in thermodynamics, the flat freezing curve indicates that the entropy of the NSF phase is almost equal to that of the solid phase.
Because the entropy of the solid phase must be small, the NSF phase has very small entropy, and appears to be a novel ordered state.

We propose that the low-entropy NSF phase is the Localized Bose-Einstein Condensation (LBEC) state.
The LBEC was first suggested by Glyde and co-workers  for $^4$He in Vycor and in Gelsil at ambient pressure~\cite{Glyde2000, Plantevin2002}.
They observed clear roton signals that prove  BEC even above the superfluid transition temperature measured by torsional oscillator technique.
The concept of LBEC is based on a hypothesis that confinement suppresses the Bose condensation temperature.
The pore size distribution of porous glass causes the spatial distribution of the BEC transition temperature.
Below bulk $T_{\lambda}$, many BECs form from large pores or intersections of pores, in which $^4$He atoms can exchange their positions frequently. 
The size of the BECs is limited to the pore size. 
The atom exchange between the BECs via the narrow regions of the pores are interrupted by the hard-core nature of $^4$He atoms. 
The whole system has therefore no global phase coherence, and does not exhibit superfluidity that can be detected by macroscopic and dynamical measurements such as a torsional oscillator. 
The localization of BECs can also be caused by disorder or randomness in the porous structure. 

It is of prime importance to determine the localized Bose condensation temperature.
We have observed a pressure minimum as previously shown in Fig.~\ref{PMinimum}.
This might be an indication of the LBEC formation, 
because bulk liquid $^4$He shows a 
pressure minimum at the $\lambda$ transition.
It might, however, be caused by  frost heaving~\cite{Hiroi1989}.
To obtain the thermodynamic evidence for the LBEC, we have been conducting a heat capacity measurement~\cite{Yamamoto2007}.

It should be noted that in the zero-temperature limit the L-S boundary tends to be 3.4 MPa, which coincides with the quantum critical pressure $P_{\mathrm c}$.
This implies that the $^4$He-nanopore system undergoes the superfluid - LBEC - solid quantum phase transition, 
although the accuracy of the pressure determination in our previous torsional oscillator study was insufficient to conclude the coincidence definitely.
Measurements near $P_{\mathrm c}$ will reveal the nature of the quantum phase transition.

In conclusion, we have determined the L-S boundary of $^4$He confined in the 2.5~nm nanopores by the pressure and thermal response.
The phase diagram shows that a novel nonsuperfluid phase exists between the superfluid and solid phases down to 0~K.
We claim that the nonsuperfluid phase is a novel LBEC state, in which macroscopic  phase coherence is destroyed by narrowness of the nanopores and random potential.

This work is supported by the Grant-in-Aid for Scientific Research on Priority Areas "Physics of Super-clean Materials" from MEXT, Japan, 
and Grant-in-Aid for Scientific Research (A) from JSPS.  
K.Y. acknowledges the support by Research Fellowships of the JSPS for Young Scientists.

\end{document}